\documentclass[aps,prl,reprint,superscriptaddress,showpacs,showkeys]{revtex4-1}%
\usepackage{amsfonts}
\usepackage{amssymb}
\usepackage{amsmath}
\usepackage{graphicx}
\usepackage{dcolumn}
\usepackage{bm}
\usepackage{units}
\usepackage{multirow}
\usepackage{CJKutf8}
\usepackage{subfigure}
\usepackage{color}%
\usepackage{url}
\usepackage[colorlinks,linkcolor=red,anchorcolor=green,citecolor=blue]{hyperref}
\providecommand{\U}[1]{\protect\rule{.1in}{.1in}}
\usepackage{array}
\newcommand{\PreserveBackslash}[1]{\let\temp=\\#1\let\\=\temp}
\newcolumntype{C}[1]{>{\PreserveBackslash\centering}p{#1}}
\newcolumntype{R}[1]{>{\PreserveBackslash\raggedleft}p{#1}}
\newcolumntype{L}[1]{>{\PreserveBackslash\raggedright}p{#1}}
\begin{document}
\title{Giant Crystal Hall Effect in Collinear Antiferromagnetic $\gamma$-FeMn}
\author{Lei Wang (\begin{CJK}{UTF8}{gbsn}王蕾\end{CJK})}
\email{wanglei.icer@xjtu.edu.cn}
\affiliation{Center for Spintronics and Quantum Systems, State Key Laboratory for Mechanical Behavior of Materials, Xi'an Jiaotong University, No.28 Xianning West Road Xi'an, Shaanxi, 710049, China}
\author{Ka Shen}
\affiliation{The Center for Advanced Quantum Studies and Department of Physics, Beijing Normal University, Beijing 100875, China}
\author{Tai Min}
\affiliation{Center for Spintronics and Quantum Systems, State Key Laboratory for Mechanical Behavior of Materials, Xi'an Jiaotong University, No.28 Xianning West Road Xi'an, Shaanxi, 710049, China}
\author{Ke Xia}
\email{kexia@csrc.ac.cn}
\affiliation{Beijing Computational Science Research Center, Beijing, 100193, China}

\date{\today}

\begin{abstract}
The spontaneous Hall effect is usually governed by three conventional mechanisms, such as the Berry curvature, skew scattering and side jump, which widely exist in ferromagnetic or antiferromagnetic materials. However, in this work, based on first principle calculations, we predict a giant crystal Hall effect (CHE) in the antiferromagnetic $\gamma$-FeMn, which can not be understood by the previous three conventional mechanisms and the Hall angle therein can be as large as 18.4\% at low temperature. Furthermore, with Boltzmann transport equation and a tight-binding model, we conclude that, the asymmetric group velocities on Fermi surface is the origin of this CHE in $\gamma$-FeMn. And with a systematic symmetry argument, we show that, this unusual effect is not dependent on specific materials but universal in any crystals with similar symmetry even without local magnetization.
\end{abstract}

\maketitle

The Hall effect and its derived anomalous or spin Hall effect are followed with great interest in the field of spintronics, which refers to a charge or spin Hall current perpendicular to the primary charge current. Basically, besides the Lorentz force from the external magnetic field, the inner contributions are mainly induced by spin-orbit interaction and time reversal symmetry breaking due to the local magnetism. For example, in the ferromagnetic metals \cite{Kondo01041962,PhysRevLett.99.086602,PhysRevB.85.220403,Grigoryan2016Scaling,PhysRevB.90.214410,PhysRevB.99.224416} and semiconductors \cite{PhysRevLett.88.207208,natm3.4,refId0}, the anomalous Hall effect has been intensively studied and the transverse anomalous Hall current is induced by the force from the spin-orbit interaction, where, similar to the spin Hall effect \cite{PhysRevLett.92.126603}, the electrons move to opposite directions determined by their spins. Thus, the anomalous Hall current can be treated as a nonzero charge current from the spin Hall effect with polarized spins. In general, the corresponding inner contributions of the Hall-like effect are usually separated to extrinsic scattering mechanisms \cite{SMIT1955877,SMIT195839,PhysRevB.2.4559} and intrinsic mechanism. The latter is independent of scattering by impurities, and is mainly governed by the Berry curvature in a momentum-space geometrical effect on Fermi energy \cite{Fang92,PhysRevLett.92.037204,PhysRevB.53.7010,PhysRevB.59.14915,PhysRevLett.93.206602,PhysRevB.76.195109}. 

Recently, with rich spin transport phenomena arising in  antiferromagnetic materials and devices \cite{PhysRevB.95.014403,2018NatNa..13..362W,2018SciA....4.3566O,2018NatCo...9..348B,Olejnik2017}, the antiferromagnetic spintronics \cite{jungwirth2018the} attracts considerable attention, including the anomalous Hall effect. Typically, the anomalous Hall effect was reported to exist in the noncollinear antiferromagnet Mn$_3$Ir \cite{PhysRevLett.112.017205}, which was confirmed in various noncollinear antiferromagnets by both theoretical and experimental works \cite{2015Natur.527..212N,PhysRevMaterials.3.044409,PhysRevB.99.104428,2017NatSR.7.515L,doi:10.1002/aelm.201800818}. Moreover, the anomalous Hall effect was also reported in collinear antiferromagnetic RuO$_2$ \cite{Smejkaleaaz8809}, which was attributed to the breaking of the time reversal symmetry by the distribution of oxygen atoms around the Ru sublattices, named crystal Hall effect (CHE). However, all the above novel anomalous Hall effect in antiferromagnetic materials are demonstrated to come from the Berry curvature, and thereby still under the control of the previous conventional regime.

\begin{figure}[tp]
	\includegraphics[width=\columnwidth]{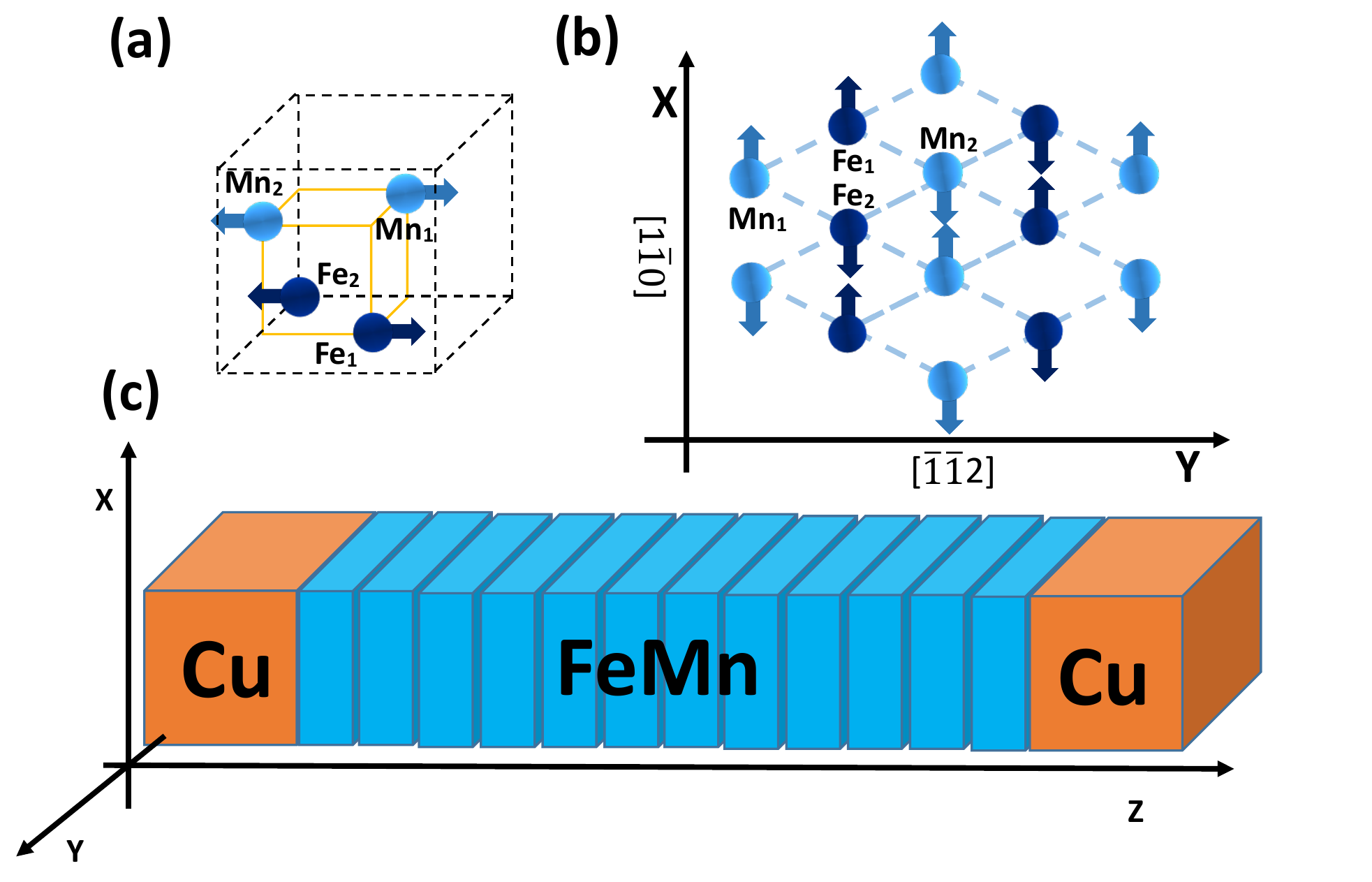}
	\caption{The transport model of the calculations. (a) The irreducible cell of the $\gamma$-FeMn with a simple cubic symmetry. (b) The in-plane ($x$-$y$ plane) spin texture for the calculations. (c) The calculated system with $\gamma$-FeMn inside the scattering region is sandwiched by two semi-infinite crystalline Cu leads, and the transport direction ($z$) is along the [111] direction of the fcc lattice. Assuming the magnetizations all parallel to $x$ axis, the generated Hall current flows along $y$ axis.}
	\label{fig1}
\end{figure}

In this work, we perform our first principle method to investigate the spontaneous Hall effect in the collinear antiferromagnetic metal, $\gamma$-FeMn. In which, the extremely large Hall angle and conductivity had been founded in low temperature. This unusual Hall effect can not be explained by previous mechanisms (e.g. Berry curvature, skew scattering and side jump), which is proved to entirely come from the crystal structure by the Boltzmann transport theory and a tight-binding model, and therefore it is a pure CHE different from that in RuO$_2$ \cite{Smejkaleaaz8809}. Moreover, with a detailed analysis of the symmetry, we conclude that this abnormal CHE is universal in any crystals with similar symmetry and not dependent on the specific material of the $\gamma$-FeMn.

The magnetic structure of $\gamma$-FeMn is shown in Fig.~\ref{fig1} (a), which has been observed by experimental measurement \cite{Bisanti_1987} and confirmed to be energetically favorable by first principle calculation \cite{PhysRevB.61.11569}. The four nonequivalent atoms locate at the corner of the fcc lattice and three neighbor face centers. In this work, the calculated magnetic moments based on the exact muffin-tin orbitals (EMTOs) \cite{andersen1995exact,vitos2007computational,emto} are $1.25~\mu_B$ and $1.92~\mu_B$ for Fe and Mn respectively, which agree well with the previous calculated value of $M_{\rm Fe}=1.4~\mu_B$ and $M_{\rm Mn}=1.9~\mu_B$ in the same crystal structure \cite{PhysRevLett.100.226602}.

For a transport system with lateral ($x$-$y$ plane) periodic boundary conditions as shown in Fig.~\ref{fig1} (c), the local charge current density from atom $R'$ to atom $R$ is calculated by an EMTO based transport code \cite{FeAHE}, which uses the Ando’s method \cite{PhysRevB.44.8017} implemented with EMTOs \cite{FeAHE,andersen1995exact,vitos2007computational,PhysRevB.100.075134,PhysRevB.102.035405} and the current operators are the same to the previous MTO base methods \cite{PhysRevB.73.064420,PhysRevB.97.214415,FeAHE,PhysRevB.102.035405,PhysRevB.65.125101, PhysRevB.77.184430,Wang.2016,PhysRevB.99.144409}, reads, 
\begin{eqnarray}
\begin{split}
{\bf j}_c(\mathbf R,\mathbf R')=({\bf R-R'}){\rm Im}\langle\Psi_R\vert\hat{\mathcal H}_{\bf R\bf R'}\vert\Psi_{R'}\rangle/\hbar .
\end{split}
\label{hc}
\end{eqnarray} 
where $|\Psi_R\rangle$ is the scattering wave functions \cite{PhysRevB.44.8017,PhysRevB.73.064420,PhysRevB.97.214415} on site $R$ and $\hat{\mathcal H}_{RR'}$ is the corresponding hopping Hamiltonian between atom $R$ and $R'$. For $\mathbf{m}\parallel x$, by projecting ${\mathbf j}_c$ to the longitudinal and transverse directions, we obtain the primary charge current $j_c^z$ and the Hall current $j_c^y$, respectively. The Hall angle is then given by $\Theta^{\rm H}_{y}=j_c^y/j_c^z$. 

The scattering geometry in our study is shown in Fig.~\ref{fig1} (c), where the scattering region $\gamma$-FeMn is sandwiched by two semi-infinite crystalline Cu leads. The transport direction ($z$-axis) is set to be along $[111]$ direction. To avoid the influence of the interfaces between the Cu leads and $\gamma$-FeMn, we use a long enough scattering region in the transport calculations, thus the Hall current in the center of the scattering region corresponds to the bulk property of the $\gamma$-FeMn. In the $x$-$y$ plane, we use 6$\times$6 lateral fcc supercells with the periodic boundary condition to generate a large ``atom box'' for disorder. Therefore, if we assumed that the equilibrium magnetizations of $\gamma$-FeMn are all in $x$ direction as shown in Fig.~\ref{fig1} (b), the conventional Hall current will flow along $y$ axis, i.e., $j_c^y$, generated by the applied current $j_c^z$. And for the sake of convenience, the fcc lattices in the Cu leads are stretched to match the $\gamma$-FeMn lattice $a_{\mathrm{FeMn}}=3.60$~\AA, in which, the transport properties extracted from the center of the scattering region are bulk properties and free from the small lattice stretch in the leads.

\begin{figure}[tp]
	\includegraphics[width=\columnwidth]{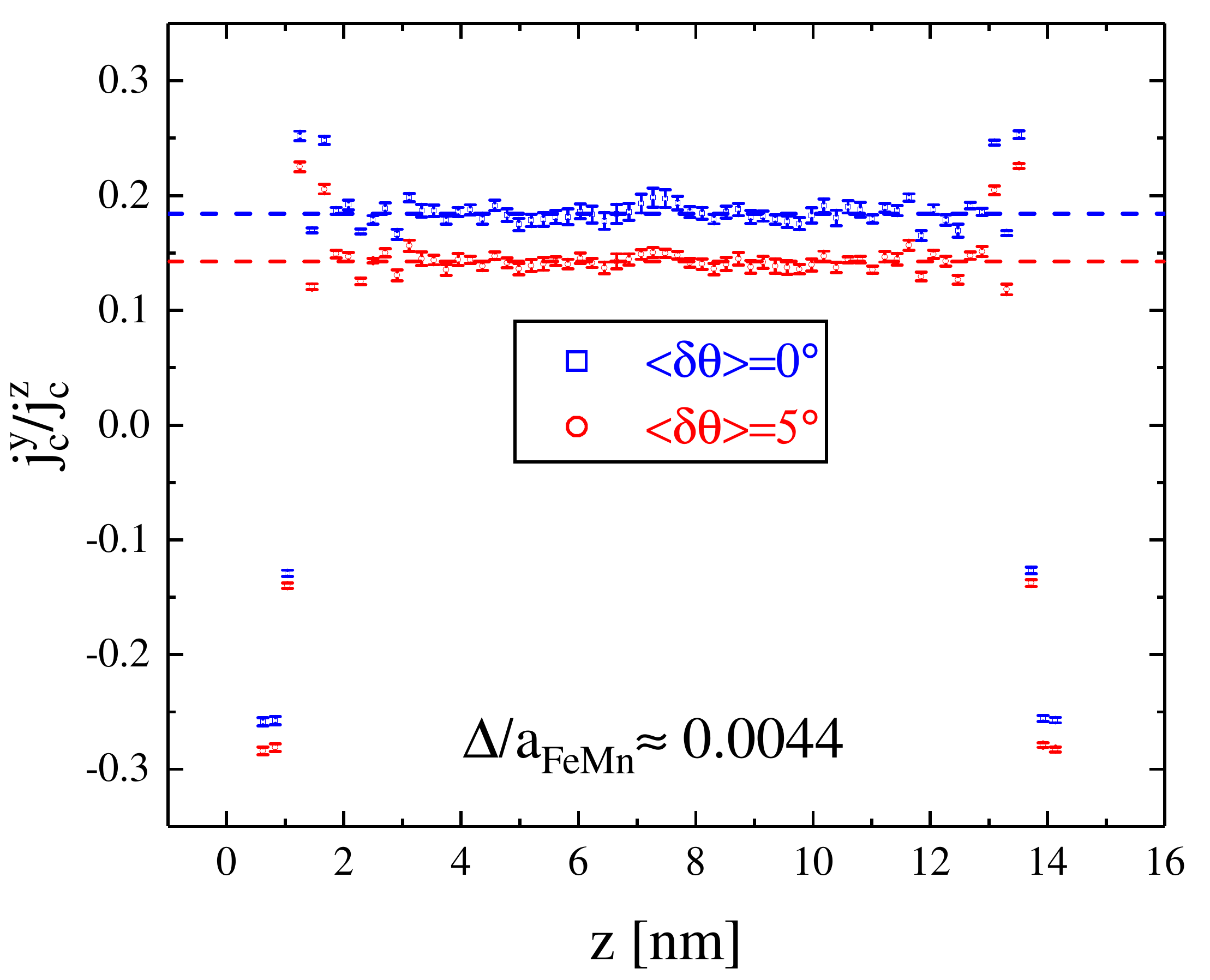}
	\caption{The calculated Hall current for a typical root mean square $\Delta/a_{\rm FeMn}\simeq0.00044$ with all magnetizations parallel to $x$-axis. Here we introduce magnetic fluctuation $\langle\delta\theta\rangle$ to simulate the magnon excited by temperature, where $\delta\theta$ is the random angle of the fluctuation of every atoms and $\langle\cdots\rangle$ represents the average over whole scattering region. The dash lines are the corresponding average value of the center of the $\gamma$-FeMn, which shows that, the Hall angles are about $\Theta^{\rm H}\simeq18.4\%$ and $\Theta^{\rm H}\simeq14.2\%$ for $\langle\delta\theta\rangle=0^{\circ}$ and $\langle\delta\theta\rangle=5^{\circ}$, respectively.}
	\label{fig2}
\end{figure}

For a perfect clean bulk material without any impurity, the electrons are only scattered by the phonon and magnon. Thus, to study the temperature dependency of the Hall current, we carry out static limit by introducing a random displacement to each atom for the calculations with phonon. The displacements are generated by a standard Gaussian distribution with a temperature dependent root-mean-square ($\Delta$) estimated by the Debye model \cite{PhysRevB.84.014412,jap1.3638694}, which has been demonstrated to be able to realize the temperature dependent resistivity, spin diffusion length, spin Hall effect, anomalous Hall effect and the results agree well with observed experiments~\cite{jap1.3638694,PhysRevB.84.014412,PhysRevB.91.220405,Wang.2016,FeAHE}. And in this work, we average 10 configurations with random displacement and discrete the lateral Brillouin zone into a 32$\times$32 mesh to converge the outputs.

Typically, for a finite root means square $\Delta/a_{\rm FeMn}\simeq0.00044$, the calculated Hall current is plotted in Fig.~\ref{fig2} by blue cubics with all magnetizations parallel to $x$-axis. We can see that, except for the sharp change around the Cu$\vert$$\gamma$-FeMn interfaces, the calculated Hall current is almost a constant, which confirms our previous expectation that the calculation inside the scattering region is all from the bulk property of the $\gamma$-FeMn. The Hall angle extracted with the red dash line is about $\Theta^{\rm H}_y\simeq18.4\%$, which is more than one order of magnitude larger than the anomalous Hall angle of Fe \cite{FeAHE} at room temperature. 

It is hard to know the magnon excitation in antiferromagnetic materials, thus we introduce a magnetic fluctuation to estimate the contributions from magnon, in which the magnetizations on every atoms are rotated from $x$-axis by a small angle $\delta\theta$ within a Gaussian distribution and a corresponding phase angles around $x$-axis within a uniform distribution in the range of $[0,2\pi]$. For a typical model with the average of the fluctuation $\langle\delta\theta\rangle=5^{\circ}$, the Hall current is calculated and the results are plotted in Fig.~\ref{fig2} by red cycles. It can be seen that, after introducing the above magnetic fluctuation, the Hall angle drops to about $\Theta^{\rm H}\simeq14.2\%$, which yields that the magnon plays an important role on the spontaneous Hall effect in antiferromagnetic material and should be studied later with more accurate model. However, as the magnon only changes the detail value of the Hall angles, and we mainly study the physical origin of this novel spontaneous Hall effect in $\gamma$-FeMn, we neglect the contributions of the magnon in the rest of the paper for simplicity and focus on the calculations at low temperature. And we know that, for $\gamma$-FeMn, the Debye temperature is $T_{D}=420$ K, together with a slightly larger Neel temperature $T_{N}=$520 K \cite{DELYAGIN201311}, thus we can easily obtain the root-mean-square \cite{jap1.3638694} for T=300 K as $\Delta/a_{\rm FeMn}\simeq0.031$, then the following calculations will be at the condition of $\Delta/a_{\rm FeMn}<0.025$, under the above considerations.

\begin{figure}[tp]
	\includegraphics[width=\columnwidth]{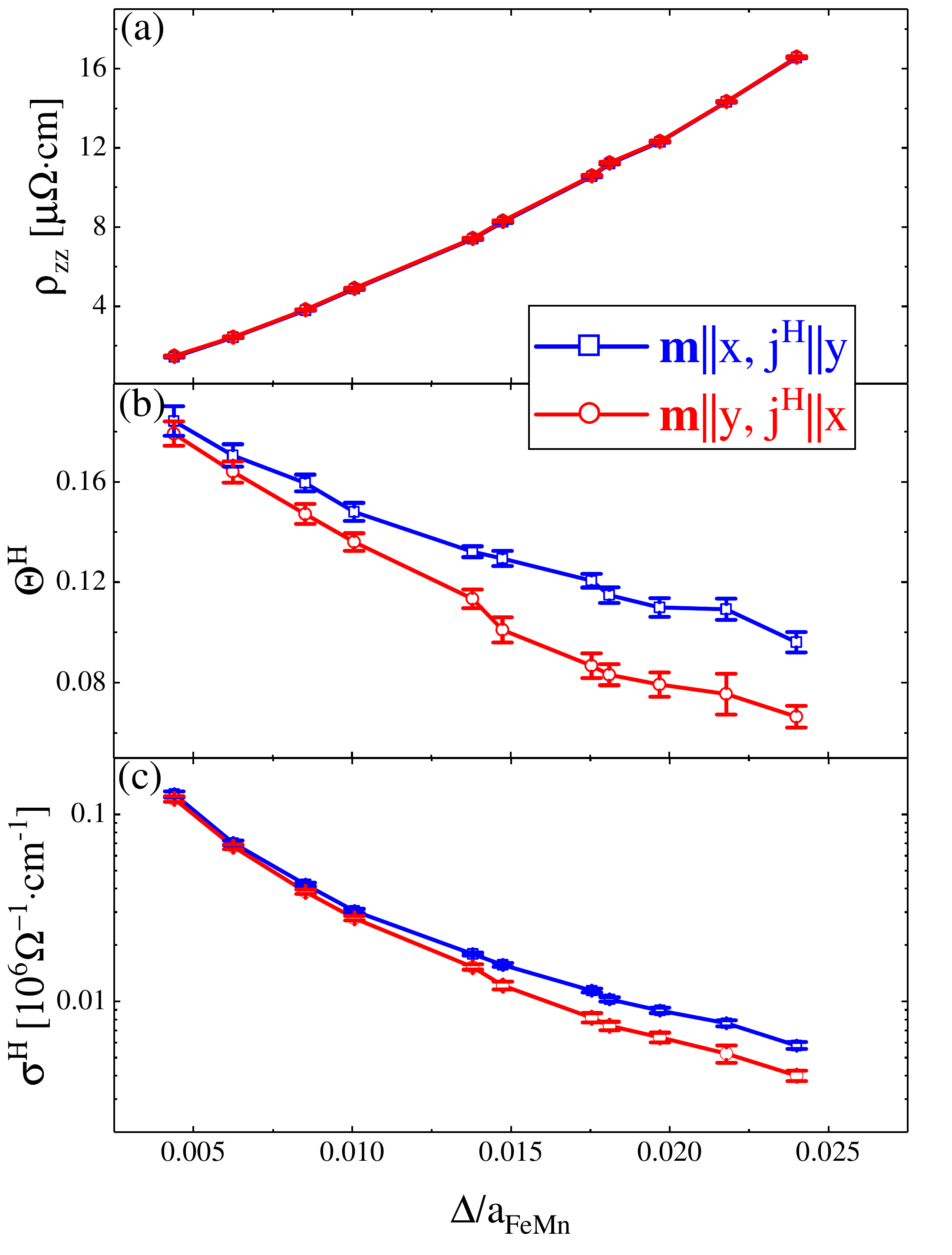}
	\caption{Calculated results for $\mathbf{m}\parallel x$ and $\mathbf{m}\parallel y$ respectively, where (a) is for the longitudinal resistivity as a function of the root-mean-square $\Delta$, (b) the corresponding Hall angles and (c) the Hall conductivities calculated by $\sigma^{\rm H}=\Theta^{\rm H}/\rho_{zz}$.}
	\label{fig3}
\end{figure}

The temperature dependence of the spontaneous Hall effect in $\gamma$-FeMn is plotted in Fig.~\ref{fig3} with $\mathbf{m}\parallel x$ and $\mathbf{m}\parallel y$, respectively. It can be seen that, the longitudinal resistivity $\rho_{zz}$ increases with increasing the temperature and the results with different direction of magnetizations are close to each other, indicating a negligible anisotropic magneto-resistance. However, as shown in Fig.~\ref{fig3} (b) and (c), the corresponding Hall angle $\Theta^{\rm H}$ and Hall conductivity by $\sigma^{\rm H}=\Theta^{\rm H}/\rho_{zz}$ decrease while increasing temperature and show a significant anisotropic effect due to the nonequivalent geometry along $x$ and $y$ direction as shown in Fig.~\ref{fig1} (b). 

\begin{figure*}[tp]
	\includegraphics[width=2.0\columnwidth]{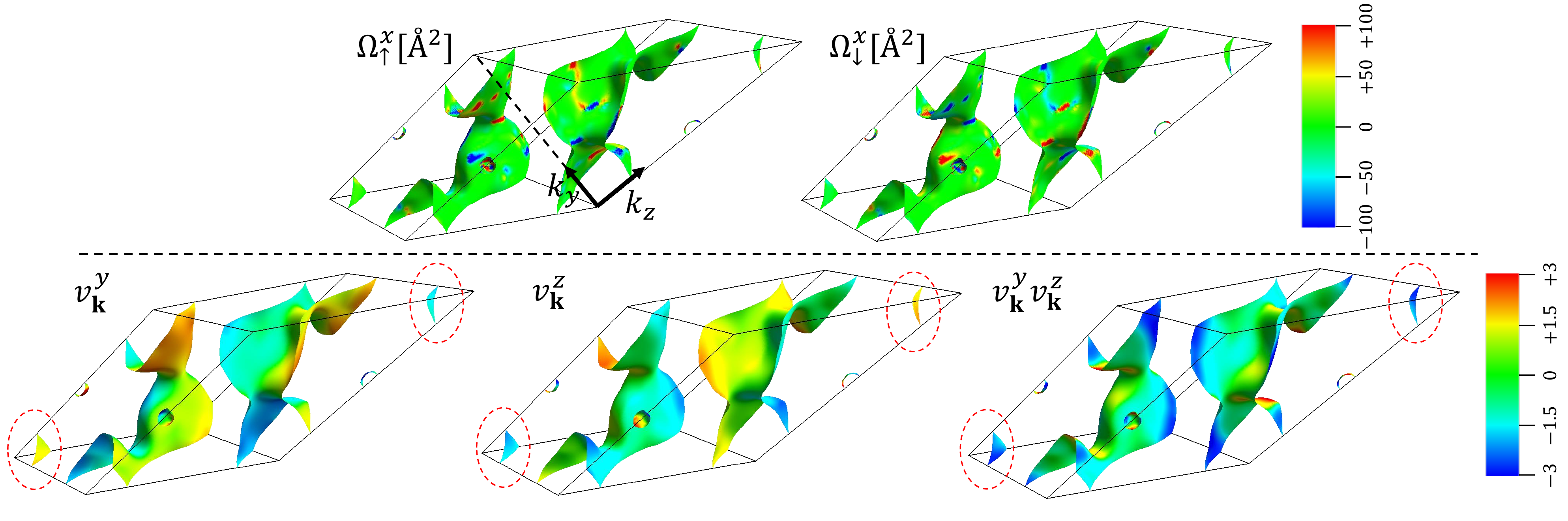}
	\caption{The calculated physical quantities on the Fermi surface of one typical band of the $\gamma$-FeMn, where $\Omega^x_{\uparrow,\downarrow}$ are the Berry curvatures corresponding to each spin and $v_{\mathbf{k}}^{y,z}$ are the group velocities in $y$ or $z$ direction, respectively.}
	\label{fig4}
\end{figure*}

Moreover, it is noticed that, the Hall angle and conductivity both reach the maximum value at clean limit as shown in Fig.~\ref{fig3}. This abnormal feature indicates that the spontaneous Hall effect in this antiferromagnetic $\gamma$-FeMn can not come from the extrinsic contributions (skew scattering and side jump), because they both induced by the scattering of the impurities according to the scaling law within the framework of multiple scattering \cite{PhysRevLett.114.217203,PhysRevB.85.220403}. To further examine the intrinsic contribution from the band structure, we calculate the Berry curvature of the $\gamma$-FeMn by the Kubo-formula derivation \cite{PhysRevLett.49.405,PhysRevLett.92.037204,PhysRevB.76.195109,si}. The calculated spin dependent Berry curvatures $\Omega^x_{\uparrow}$ and $\Omega^x_{\downarrow}$ on the Fermi surface of one typical band are plotted in Fig.~\ref{fig4}. As seen that, $\Omega^x_{\uparrow}=-\Omega^x_{\downarrow}$ for all $\mathbf{k}$ points in the whole Fermi surface, meaning that the degenerate spin states give opposite Berry curvature. Since all bands in $\gamma$-FeMn are spin degenerate due to the antiferromagnetic nature, the total Berry curvature will be zero, and therefore, there is no net intrinsic contribution to the spontaneous Hall effect.

On top of the above discussions, we can conclude that this unusual spontaneous Hall effect in $\gamma$-FeMn is not governed by the extrinsic skew scattering, side jump and Berry curvature. There should be a new mechanism that contributes to the spontaneous Hall effect and dominates in this antiferromagnetic $\gamma$-FeMn. In this sense, we turn to the Boltzmann transport equation to reveal the fundamental physical picture of the spontaneous Hall effect, in which the Hall current for a specific energy $\epsilon_{\mathbf{k}}$ can be obtained as \cite{si},
\begin{eqnarray}
	j_c^y=\tau e^2\vert\mathbf{E}\vert
	\sum_{\mathbf{k}}v_{\mathbf{k}}^y v_{\mathbf{k}}^z \delta(\epsilon_{\mathbf{k}}-\epsilon_F).
	\label{iy}
\end{eqnarray}
where $\tau$ is the relaxation time of the electrons, $e$ the electron charge, $\mathbf{E}$ the electric field, $v_{\mathbf{k}}^y$ ($v_{\mathbf{k}}^z$) the corresponding group velocities along $y$ ($z$) axis, and $\epsilon_F$ the Fermi energy.

\begin{figure}[tp]
	\includegraphics[width=\columnwidth]{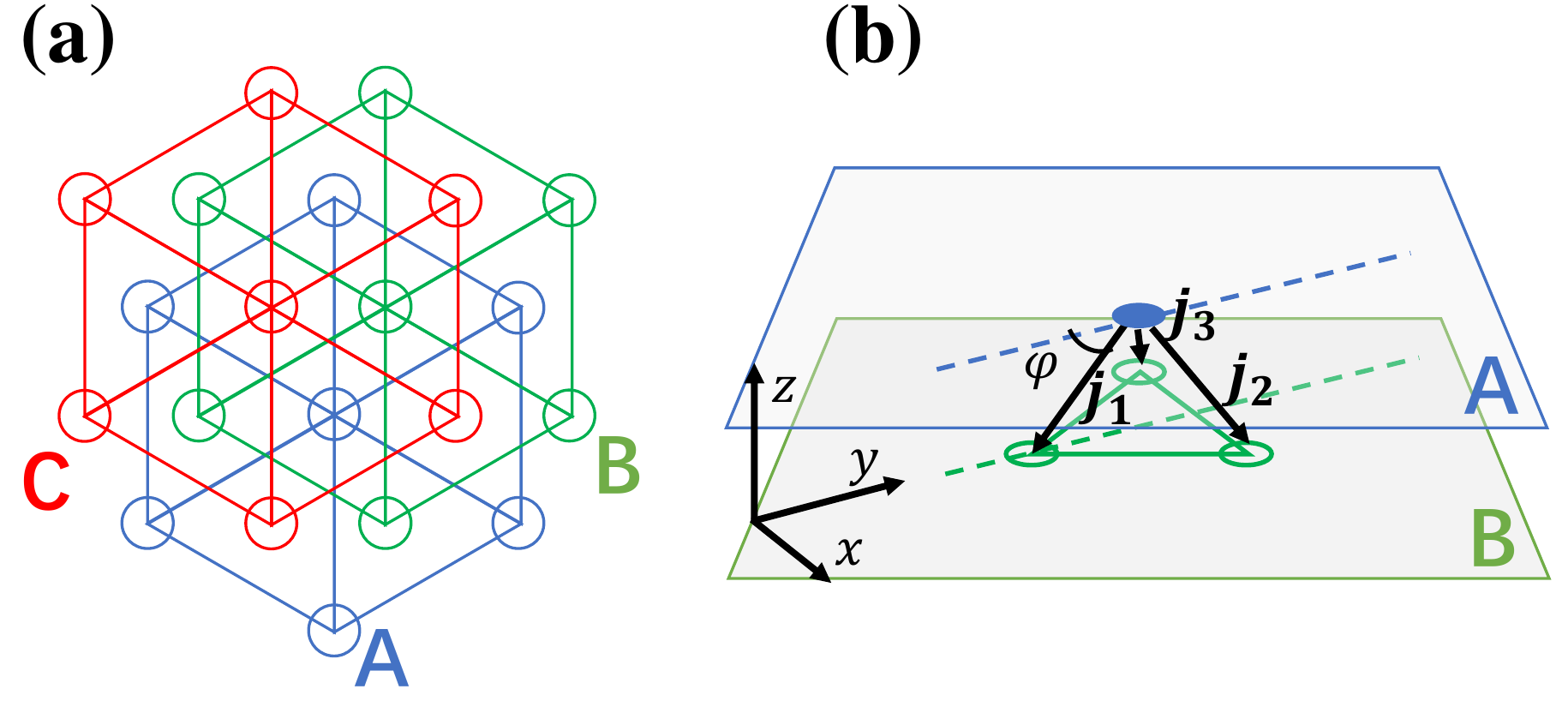}
	\caption{The sketch of the microcosmic origin of the CHE in the antiferromagnetic $\gamma$-FeMn, (a) the corresponding relation of the A-B-C plane along [111] direction of the fcc structure, (b) the diagram of the local charge current between neighbor atoms in A-B planes.}
	\label{fig5}
\end{figure}

Therefore, the detail symmetry of the $v_{\mathbf{k}}^y$ and $v_{\mathbf{k}}^z$ on Fermi surface will determine the appearance of the Hall current. Thus, we calculate the group velocities on Fermi surface of the $\gamma$-FeMn as shown in Fig.~\ref{fig4}. It can be seen that, especially inside the dash red circles, $v_{\mathbf{k}}^y$ and $v_{\mathbf{k}}^z$ are always with different signs, thus the corresponding $v_{\mathbf{k}}^yv_{\mathbf{k}}^z$ leads to a non-zero Hall current according to Eq.~(\ref{iy}).

For a deep understanding of the Hall current in the present case, we go to the atomic level to analyze the hopping through A-B-C plane of the fcc structure along the $z$-axis as shown in Fig.~\ref{fig5} (a), in which the injected charge current is perpendicular to all the planes. We can see that, all atoms are located at the center of the triangles constructed by the three neighbor atoms in the near planes. Considering only the neighbor atoms, as shown in Fig.~\ref{fig5} (b), the injected charge current flows from the atom in A plane to three atoms in B plane, which are marked as $j_{1,2,3}$, and have the same angle $\varphi$ with A plane. Following these considerations, we can write down the currents as,
\begin{eqnarray}
	\begin{split}
		j_c^z&=-\left(j_1+j_2+j_3\right)\sin\varphi \\
		j_c^y&=\left(-j_1+\frac{1}{2}j_2+\frac{1}{2}j_3\right)\cos\varphi
	\end{split}
	\label{strcur}
\end{eqnarray}
where $j_c^z$ is the injected charge current and $j_c^y$ the induced Hall current. It can be seen that, in $\gamma$-FeMn, the $j_{1,2,3}$ connected three atoms in B plane have not only different type of atoms, but also anti-parallel magnetizations as shown in Fig.~\ref{fig1} (b). Thus, the above anisotropic effect will end up with nonequivalent $j_{1,2,3}$, and then nonzero Hall current $j_c^y$ appears using Eq.~(\ref{strcur}), which is the microcosmic origin of the crystal structure induced Hall effect in $\gamma$-FeMn. As this spontaneous Hall effect is dominated by the symmetry of the crystal structure, we name it as the CHE to distinguish it from the conventional anomalous Hall effect. To verify our assumptions above, we choose a typical Fe atom in the middle of the scattering region as the central atom in plane A, and calculated $j_{1,2,3}$ from first principle with Eq.~(\ref{hc}). The numerical results show that $j_1:j_2:j_3\simeq1.33:1.73:2.28$, which supports our theory.
 
In summery, we predict a giant CHE in the collinear antiferromagnetic $\gamma$-FeMn using the first principle method. Our results show that, the Hall angle at low temperature can be one order of magnitude larger than that in the ordinary ferromagnetic materials, such as Fe. Moreover, this novel CHE is beyond the current conventional anomalous Hall effect, which is proved to come from the asymmetric group velocities on Fermi surface by the Boltzmann transport theory and a tight-binding model. And with detail symmetry argument, this structure induced CHE should be universal and exist in any materials with similar symmetry.
\begin{acknowledgments}
This work was supported the National Natural Science Foundation of China (grant No. 11804266). K. X. is supported by the National Key Research and Development Program of China (grant Nos. 2017YFA0303300 and 2018YFB0407601), the National Natural Science Foundation of China (grant Nos. 61774017, 11734004 and 21421003), and NSAF (grant No. U1930402). T. M. is supported by the National Key Research and Development Program of China (grant Nos. 2018YFB0407600, 2017YFA0206202 and 2016YFA0300702) and Shaanxi Province Science and Technology Innovation Project (grant 2019TSLGY08-04). K. S. is supported by the National Natural Science Foundation of China (grant No. 11974047).
\end{acknowledgments}

\bibliographystyle{apsrev4-1}
\bibliography{icer_FeMn}

\begin{thebibliography}{71}
\bibitem [{not()}]{notex}%
  \BibitemOpen
  \href@noop {} {}\bibinfo {note} {Note that the direction of the magnetization
  can be $-x$ in antiferromagnetic case. Here we still assume that the
  magnetizations are parallel to $x$-axis and treat the location of the
  corresponding atoms as a different type point in the point group analysis to
  deal with the symmetry argument.}\BibitemShut {Stop}%
\end{thebibliography}

\clearpage
\section*{supplementary materials}
\subsection{A. Berry curvature calculations}

Technically, we first calculate the electronic structure of the $\gamma$-FeMn with the same crystal structure as described in the main text by the VASP (Vienna ab-initio simulation package) code \cite{PhysRevB.47.558,PhysRevB.54.11169}, which is based on density function theory (DFT) and the generalized gradient approximation (GGA) with an interpolation formula according to Vosko, Wilk, and Nusair \cite{1980CaJPh} and a plane-wave basis set within the framework of the projector augmented wave (PAW) method \cite{PhysRevB.50.17953,PhysRevB.59.1758}. In detail, the cut-off energy for the basis is 500 eV, and the convergence criterion for the electron density self-consistency cycles is $10^{-6}$ eV. In the Brillouin zone, we sample ($6\times6\times12$) k-point grids using the Monkhorst-Pack scheme \cite{PhysRevB.13.5188} to converge the outputs. Also the spin-orbit coupling is introduced for convenient of the study of Berry curvature.

To calculate the Hall conductivity from the contribution of the Berry curvature, the well-known formula \cite{PhysRevLett.92.037204,PhysRevB.76.195109} is used, and typically when the magnetization parallel to $x$ direction, reads,
\begin{eqnarray}
	\sigma^{\rm H}_{intr}=-\frac{e^2}{\hbar}\int_{BZ}\frac{d^3k}{(2\pi)^3}\Omega^x(\mathbf{k})
\end{eqnarray}
where $\hbar$ is the Planck constant, ``$BZ$'' represents the integration over the total Brillouin zone, $\mathbf{k}$ is the wave vector, and $\Omega^x(\mathbf{k})$ is the sum of the Berry curvatures over the occupied bands for each $\mathbf{k}$:
\begin{eqnarray}
	\Omega^x(\mathbf{k})=\sum_n f_n \Omega_n^x(\mathbf{k})
\end{eqnarray}
where $n$ is the quantum number of all the occupied bands and $f_n$ is the corresponding equilibrium Fermi-Dirac distribution, and the Berry curvature arises from the Kubo-formula derivation \cite{PhysRevLett.49.405}, reads,
\begin{eqnarray}
	\Omega_n^x(\mathbf{k})=-\sum_{n'\neq n}\frac{2\mathrm{Im}\langle\psi_{n\mathbf{k}}\vert v_y \vert\psi_{n'\mathbf{k}}\rangle \langle\psi_{n'\mathbf{k}}\vert v_z \vert\psi_{n\mathbf{k}}\rangle}{(\omega_{n'}-\omega_n)^2}
\end{eqnarray}
where the energy of each band $E_n=\hbar\omega_n$, $v_{y,z}$ are velocity operators and $\psi$ is the wave function.

The above formula had already been generated in the open source code ``WANNIER90'' \cite{Pizzi_2020} and ``Wannier Berri'' \cite{wannier-berri1,wannier-berri2} with the maximally localized generalized Wannier functions (MLWFs) \cite{PhysRevB.56.12847,PhysRevB.65.035109,RevModPhys.84.1419} which can connected to the previous electronic structure from VASP code conveniently. In addition, we use a three-dimensional $\mathbf{k}$ mesh in the total Brillouin zone with the spacing of $\mathbf{k}$-points being $\Delta\mathbf{k}\simeq\frac{2\pi}{\textit{Len}}$, where $\textit{Len}=200$, typically. Moreover, to make the calculation more precise around the typical $\mathbf{k}$ points with major contribution to the Berry curvature, the adaptive recursive refinement algorithm \cite{wannier-berri1} is used, and we calculate 30 iterations to make sure the Berry curvature calculations converged. 

\subsection{B. Boltzmann transport equation}
To reveal the fundamental physical picture of the crystal Hall effect (CHE), we start from the Boltzmann transport equation to reproduce the Hall effect, reads,
\begin{eqnarray}
	\partial_t f_{\mathbf{k}}+\mathbf{v}_{\mathbf{k}}\cdot\nabla f_{\mathbf{k}}+e\mathbf{E}\cdot\nabla_{\mathbf{k}} f=-\frac{\delta f}{\tau}
\end{eqnarray}
where $f_{\mathbf{k}}$ is the Fermi-Dirac distribution of the electrons with a specific wave vector $\mathbf{k}$, $\mathbf{v}_{\mathbf{k}}$ is the corresponding group velocity, $e$ is the electron charge, $\mathbf{E}$ is the electric field and $\tau$ is the relaxation time of the electrons. In a stable and homogeneous system, $\partial_t f_{\mathbf{k}}=0$ and $\nabla f_{\mathbf{k}}=0$, thus, we have
\begin{eqnarray}
	-\tau e\mathbf{E}\cdot\nabla_{\mathbf{k}} f=\delta f.
\end{eqnarray}
And described in the main text, the primary charge current is along $z$ direction ($\mathbf{E}\parallel z$) and the induced Hall current is in $y$ direction, therefore, the Hall current is written as,
\begin{eqnarray}
	\begin{split}
		j_c^y&=\sum_{\mathbf{k}}e v_{\mathbf{k}}^y\delta f \\
		&=-\tau e^2\vert\mathbf{E}\vert
		\sum_{\mathbf{k}}v_{\mathbf{k}}^y(\partial_{\epsilon_{\mathbf{k}}} f)(\partial_{k_z} \epsilon_{\mathbf{k}}).
	\end{split}
\end{eqnarray}
At low temperature, $\partial_{\epsilon_{\mathbf{k}}} f=-\delta(\epsilon_{\mathbf{k}}-\epsilon_F)$ with $\epsilon_F$ is the Fermi energy, thus the above formula becomes 
\begin{eqnarray}
	j_c^y=\tau e^2\vert\mathbf{E}\vert
	\sum_{\mathbf{k}}v_{\mathbf{k}}^y v_{\mathbf{k}}^z \delta(\epsilon_{\mathbf{k}}-\epsilon_F),
	\label{iy}
\end{eqnarray}
which can be used to analysis the CHE based on the group velocity on the Fermi surface.

\subsection{C. Symmetry argument}
The key point of the predicted novel CHE is that whether the in plane symmetry of the [111] plane of the $\gamma$-FeMn allows an Hall current arising from the primary charge current $j_c^z$ accordingly. Thus we carry out the symmetry analysis on the resistivity tensor using the well studied symmetry argument in the anisotropic magneto-resistance investigations \cite{Mcguire1975Anisotropic,MR0180308}, where we can define a vector in a crystallographic axes with the cosines to describe the unit vector of the magnetization, e.g. $\hat{\alpha}=(\alpha_1,\alpha_2,\alpha_3)$. And in this regime, we have the relation $\mathbf{E}=\hat{\rho}(\hat{\alpha})\mathbf{J}$, where $\mathbf{E}$ is the electric field, $\mathbf{J}$ is the current and $\hat{\rho}(\hat{\alpha})$ is the resistivity tensor, which can be expanded in a MacLaurin's series with the Einstein summation convention \cite{Mcguire1975Anisotropic}
\begin{eqnarray}
	\rho_{ij}(\hat{\alpha})=a_{ij}+a_{ijk}\alpha_k+a_{ijkl}\alpha_k\alpha_l+\cdots
\end{eqnarray} 
By expressing the current as $\mathbf{J}=\vert\mathbf{J}\vert\hat{\beta}$ with $\hat{\beta}=(\beta_1,\beta_2,\beta_3)$, the Hall resistivity along $\hat{\gamma}=(\gamma_1,\gamma_2,\gamma_3)$ direction will be
\begin{eqnarray}
	\rho^{\rm H}(\hat{\alpha},\hat{\beta},\hat{\gamma})=\frac{\hat{\gamma}\cdot\mathbf{E}}{\vert\mathbf{J}\vert}=\hat{\gamma}\cdot[\hat{\rho}(\hat{\alpha})\cdot\hat{\beta}]
\end{eqnarray}

\begin{figure}[b]
	\includegraphics[width=\columnwidth]{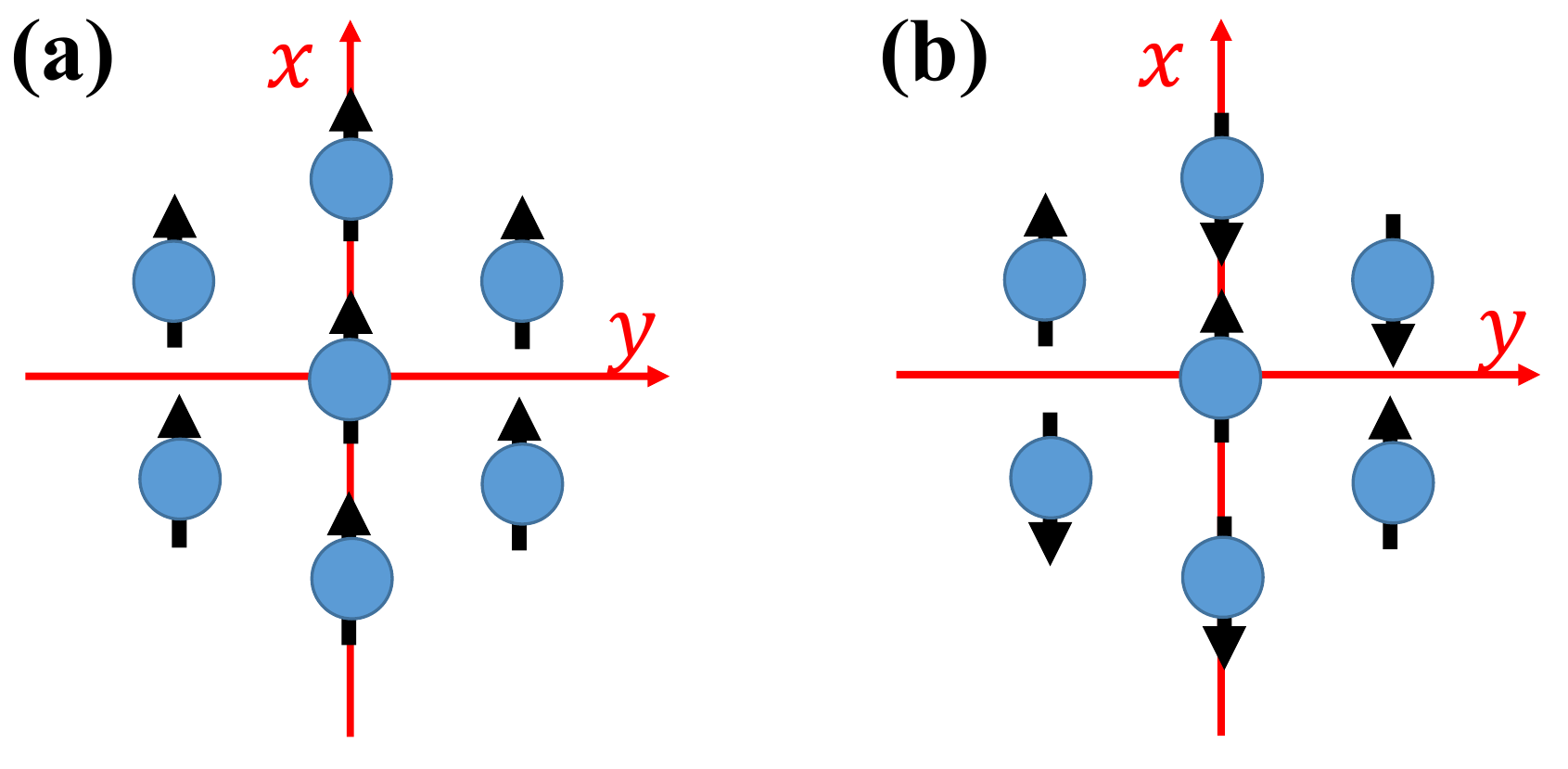}
	\caption{The sketch of the [111] plane of the fcc structure with the magnetization parallel/anti-parallel to $x$-axis and charge current in $z$-axis. Here (a) is for a normal fcc ferromagnetic material, (b) the antiferromagnetic $\gamma$-FeMn.}
	\label{fig5}
\end{figure}

In our configurations, the charge current is applied in $z$-axis and the magnetic moments are all parallel to the $x$-axis \cite{notex}, therefore the Hall current will be along $y$-axis as shown in Fig.~\ref{fig5}. Then it is obvious that $\hat{\alpha}=(1,0,0)$, $\hat{\beta}=(0,0,1)$ and $\hat{\gamma}=(0,1,0)$. And the Hall resistivity will be
\begin{eqnarray}
	\rho^{\rm H}(\hat{\alpha},\hat{\beta},\hat{\gamma})=\rho_{23}.
\end{eqnarray}
For simplicity, we expand the $\hat{\rho}(\hat{\alpha})$ to first order
\begin{eqnarray}
	\rho_{23}=a_{23}+\sum_k a_{23k}\alpha_k=a_{23}+a_{231}.
\end{eqnarray}

It is known that, the coefficients are related to the symmetry, so if $t_{ij}$ is the element of a transformation matrix that leaves the crystal unchanged, then \cite{MR0180308}
\begin{eqnarray}
	a_{mn}=\sum_{ij}t_{mi}t_{nj}a_{ij}
\end{eqnarray}
and 
\begin{eqnarray}
	a_{mnl}=\vert \hat{t}\vert\sum_{ijk} t_{mi}t_{nj}t_{lk}a_{ijk}
\end{eqnarray}
where $\vert \hat{t}\vert=1$ stands for the determinant of the transformation matrix $\hat{t}$. Also, we know that after transformation by the $\hat{t}$ matrix, everything should be the same, which indicates that $a_{mn}=a_{ij}$ and $a_{mnl}=a_{ijk}$. On top of the above equations, we can simply analyze the in-plane symmetry of one single plane since each atomic layer along the $z$-axis (fcc [111] direction) can be handled similarly.

For an ordinary ferromagnetic material with fcc structure as shown in Fig.~\ref{fig5} (a), there is a $60^{\circ}$ rotation symmetry around $z$ axis ($C_{6v}$), described by
\begin{eqnarray}
	\hat{t}=\left(
	\begin{array}{ccc}
		\frac{1}{2} & -\frac{\sqrt{3}}{2} & 0 \\
		\frac{\sqrt{3}}{2} &  \frac{1}{2} & 0 \\
		0 &  0 & 1
	\end{array}
	\right)
\end{eqnarray}
leading to $a_{23}=\sum_{ij}t_{2i}t_{3j}a_{ij}=\sqrt{3}a_{13}$ and $a_{13}=\sum_{ij}t_{1i}t_{3j}a_{ij}=-\sqrt{3}a_{23}$. Thus, $a_{23}=0$.

Similarly,
\begin{eqnarray}
	a_{231}=\vert \hat{t}\vert\sum_{ijk} t_{2i}t_{3j}t_{1k}a_{ijk}=\sum_{ijk} t_{2i}t_{3j}t_{1k}a_{ijk}
\end{eqnarray}
with few algebra process, it gives $a_{231}=-2a_{132}$. We already know that after the transformation by $\hat{t}$ matrix, the coefficients $a$ should be the same, therefore, $a_{231}=-2a_{132}=a_{132}=0$. And finally we obtain $\rho^{H}(\hat{\alpha})=0$. Thus there is no CHE from structure in the conventional fcc ferromagnetic materials, implying that the anomalous Hall effect should originate from spin-orbit interaction only.

For $\gamma$-FeMn case, the in-plane symmetry is shown in Fig.~\ref{fig5} (b), and due to the antiferromagnetic property, it only has a $180^{\circ}$ rotation symmetry around $z$ axis ($C_{2v}$), described by
\begin{eqnarray}
	\hat{t}=\left(
	\begin{array}{ccc}
		-1 & 0 & 0 \\
		0  & -1 & 0 \\
		0 &  0 & 1
	\end{array}
	\right),
\end{eqnarray}
using a similar method, we will have $a_{23}=0$. 

But interestingly, 
\begin{eqnarray}
	a_{231}=\vert \hat{t}\vert\sum_{ijk} t_{2i}t_{3j}t_{1k}a_{ijk}=\sum_{ijk} t_{2i}t_{3j}t_{1k}a_{ijk}
\end{eqnarray}
which ends up with a relation that $a_{231}=a_{231}$, means that $a_{231}$ could be any value within the symmetry requirement. Thus, the in plane symmetry of the $\gamma$-FeMn allows a Hall current generated in $y$-axis, and the Hall resistivity from the structure will be
\begin{eqnarray}
	\rho^{\rm H}(\hat{\alpha})=a_{231}.
\end{eqnarray}

\begin{figure}
	\includegraphics[width=\columnwidth]{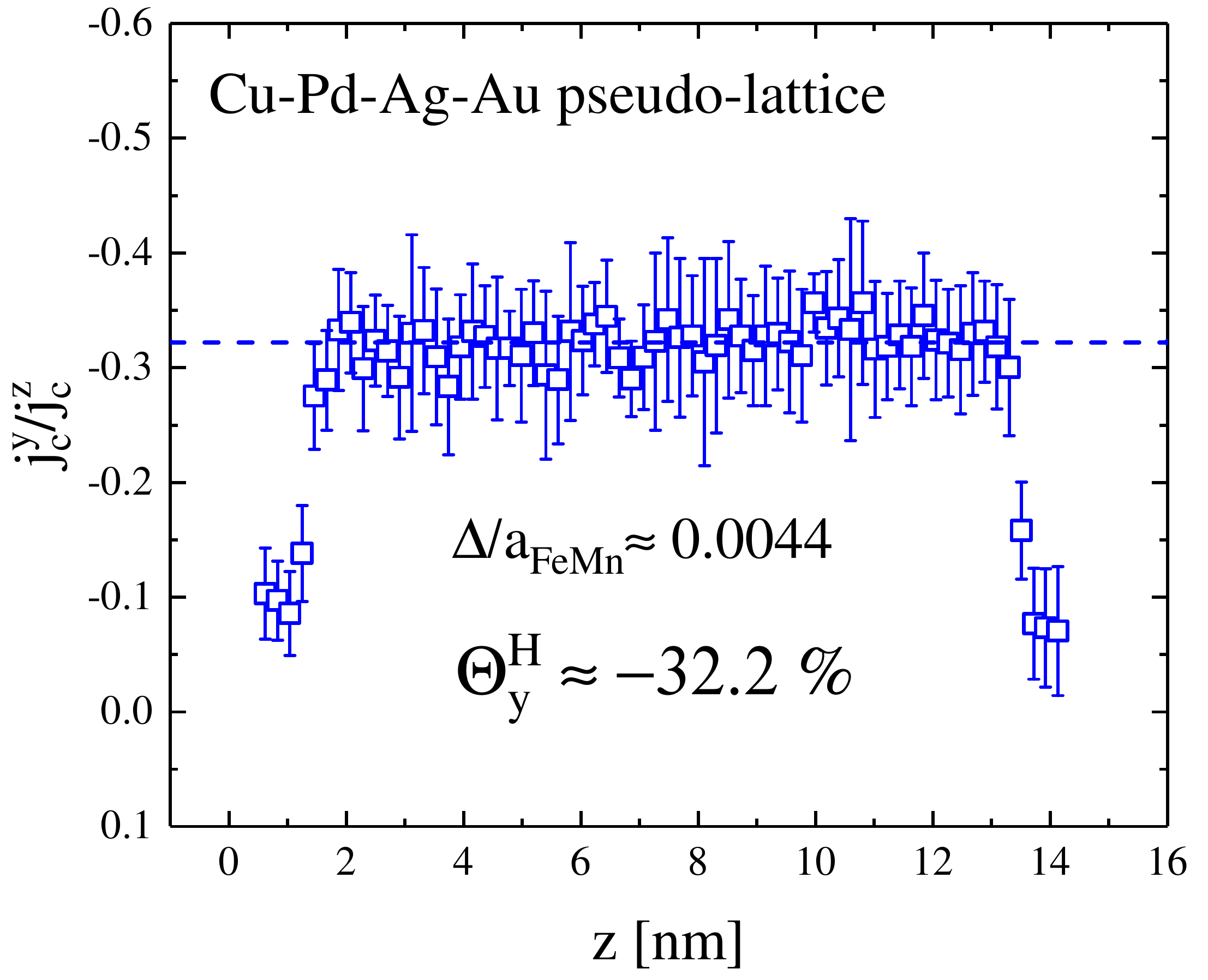}
	\caption{The normalized Hall current $j_c^y/j_c^z$ in a nonmagnetic pseudo-lattice, where the potentials of the four nonequivalent atoms in $\gamma$-FeMn are replaced by the  potentials of bulk Cu, Pd, Ag, and Au with the same lattice constant, respectively. }
	\label{fig6}
\end{figure}

Furthermore, according to the above analysis, the antiferromagnetic $\gamma$-FeMn is not necessary for the induced CHE. And this symmetry broken induced CHE exists in any materials with similar symmetry, such as permalloy \cite{PyAHE} and the materials even without any local magnetization. To confirm such conclusion, we construct a nonmagnetic pseudo-lattice to calculate the CHE, where, the crystal lattice of $\gamma$-FeMn is kept the same, and the corresponding four nonequivalent atoms are replaced by nonmagnetic Cu, Pd, Ag, and Au, respectively. The calculated results are plotted in Fig.~\ref{fig6}, in which, it can be seen that, there exists giant CHE and the corresponding Hall angle can be as large as $\Theta^{\rm H}_y\simeq-32.3\%$.

\end{document}